\documentclass[fp,twocolumn]{jpsj3}
\usepackage{txfonts}
\usepackage{bm}
\usepackage{color}
\usepackage{ulem}
\usepackage[dvipdfmx]{graphicx}
\usepackage{dcolumn}
\usepackage{mathrsfs}
\usepackage{ulem}
\usepackage{xcolor}
\usepackage{multirow}
\title{Field-Induced Criticality in YbCu$_4$Au}

\author{Takanori Taniguchi$^1$\thanks{takanori.taniguchi.d3@tohoku.ac.jp}, Kotaro Osato$^{1,2}$, Hirotaka Okabe$^1$, Takafumi Kitazawa$^{1,2}$, Masahiro Kawamata$^{1,2}$, Shota Hashimoto$^{1,2}$, Yoichi Ikeda$^1$, Yusuke Nambu$^{1,3,4}$, Dita Puspita Sari$^{5,6}$, Isao Watanabe$^6$, Jumpei G. Nakamura$^7$, Akihiro Koda$^7$, Jun Gouchi$^8$, Yoshiya Uwatoko$^8$, Shunichiro Kittaka$^9$, Toshiro Sakakibara$^8$, Masaichiro Mizumaki$^{10}$, Naomi Kawamura$^{11}$, Takayoshi Yamanaka$^1$, Koichi Hiraki$^{12}$, Takahiko Sasaki$^1$, and Masaki Fujita$^1$}
\inst{$^1$Institute for Materials Research, Tohoku University, Sendai 980-8577, Japan \\
$^2$Department of Physics, Tohoku University, Sendai 980-8578, Japan \\
$^3$Organization for Advanced Studies, Tohoku University, Sendai 980-8577, Japan\\
$^4$FOREST, Japan Science and Technology Agency, Kawaguchi, Saitama 332-0012, Japan\\
$^5$Innovative Global Program, College of Engineering, Shibaura Institute of Technology, Saitama 337-8570, Japan\\
$^6$Nuclear Structure Research Group, Nishina Center for Accelerator-Based Science, RIKEN, Wako, Saitama 351-0198, Japan\\
$^7$Institute of Materials Structure Science, High-Energy Accelerator Research Organization, Tsukuba 305-0801, Japan \\
$^8$The Institute for Solid State Physics, The University of Tokyo, Kashiwa, Chiba 277-8581, Japan\\
$^9$Department of Physics, Faculty of Science and Engineering, Chuo University, Tokyo, 112-8551, Japan\\
$^{10}$Faculty of Science, Kumamoto University, 2-39-1 Kurokami Chuou-ku, Kumamoto 860-8555, Japan\\
$^{11}$Japan Synchrotron Radiation Research Institute, Sayo, Hyogo 679-5148, Japan\\
$^{12}$Integrated Center for Science and Humanities, Fukushima Medical University, Fukushima 960-1295, Japan\\
}

\abst{
YbCu$_4$Au is a unique material exhibiting multiple quantum fluctuations simultaneously. In this study, we investigated the field-induced criticality in YbCu$_4$Au, based on comprehensive micro and macro measurements, including powder X-ray diffraction (XRD), neutron powder diffraction (NPD), nuclear magnetic resonance, magnetization, resistivity, specific heat, muon spin rotation relaxation ($\mu$SR), and X-ray absorption spectroscopy (XAS). Single crystals of YbCu$_4$Au were grown, and their crystal structure was determined using XRD, and NPD measurements. Magnetic successive transitions were observed below 1 T by specific heat, resistivity, NPD, and $\mu$SR measurements. XAS measurements further indicate that the valence of Yb ions ($+2.93$) remained unchanged above 2 T. Moreover, the change in quadrupole frequency observed in the previous study is attributable to the electric quadrupole, as the expected value of the electric quadrupole was finite under magnetic fields [S. Wada $et$ $al$., $Journal$ $of$ $Physics$: $Condensed$ $Matter$, $\mathbf{20}$, 175201 (2008).]. These experimental results suggest that YbCu$_4$Au exhibited bicritical behavior near 1 T, arising from the competition between RKKY interaction, accounting for the magnetic phases, and the Zeeman effect.}

\begin{document}
\maketitle

\section{Introduction}
The studies of quantum criticality have been attracted attention in solid-state physics for the exploration of exotic physical properties. 
Establishing the connection between ordered phases, which drive quantum critical phenomena, and quantum criticality is crucial for understanding this subject. 
A potent experimental strategy for investigating this connection involves manipulating the interactions originating from ordered phases using external parameters, such as magnetic fields~\cite{Doniach_1977, Gegenwart_2008, Si_2010}. 
Among the primary interactions that lead to magnetically ordered phases in metallic magnetic materials, the Ruderman$\--$Kittel$\--$Kasuya$\--$Yosida (RKKY) interaction is particularly significant. This interaction is closely associated with the Fermi surface, and adjusting external parameters enables the RKKY interaction to undergo various transitions~\cite{Mignod_1997, Taniguchi_2023}.

A small energy scale of $f$ electrons presents an ideal scenario for investigating quantum criticality, as it enables the experimental modulation of certain interactions via external parameters. Close to the quantum critical point (QCP), where the magnetic transition temperature approaches absolute zero, the competition between RKKY interaction and Kondo effect has been reported to lead to anomalous physical properties, such as unconventional superconductivity and spin liquids~\cite{Gegenwart_2008, Si_2010}. These phenomena are theoretically underpinned by the Doniach phase diagram~\cite{Doniach_1977} and can be attributed to antiferromagnetic spin fluctuations~\cite{Moriya_1985}.
Conversely, in U-based compounds with ferromagnetic spin fluctuations, a phase diagram with a tricritical point has been proposed~\cite{Tokunaga_2015,Taufour_2016,Nakamura_2017, Aoki_2019}. This phase diagram is three-dimensional, with magnetic field components along two crystal axes and temperature. In URhGe~\cite{Tokunaga_2015,Taufour_2016,Nakamura_2017, Aoki_2019}, the $B_{\rm b}$$\--$$B_{\rm c}$$\--$$T$ phase diagram, involving the magnetic field components along the b-axis, c-axis, and temperature, has been reported, where a tricritical point exists at the junction of two second-order phase transition lines and one first-order phase transition line. The ferromagnetic superconductivity discovered in these uranium-based compounds has been extensively discussed within this framework, and it is highly likely that Cooper pairs are in a spin-triplet state~\cite{Aoki_2019, Mineev_2015, Lewin_2023}. To investigate points such as the tricritical point, where some phase transition lines merge, is an excellent strategy to discover exotic physical properties, as the spin fluctuations are enhanced. Therefore, the search for candidate materials is ongoing.

Quantum critical phenomena driven by degrees of freedom beyond spin have garnered increasing attention, with valence fluctuations identified as a research hotspot. Theoretical analyses have shown that heavy-fermion systems, characterized by a pronounced Kondo effect, may reach a valence-based QCP when the repulsion between conduction and $f$ electrons competes with the kinetic energy of the $f$ electrons~\cite{Watanabe_2010}. However, materials that exhibit valence quantum criticality cannot easily be identified because the number of candidate materials is also limited.

The YbCu$_4T$ family, where $T$ represents a transition metal, is an exemplary system for investigating valence quantum criticality, as demonstrated by its exhibited valence transitions~\cite{Nakamura_1990,Kojima_1990,Yamaoka_2023}, states of valence fluctuation~\cite{Anzai_2019}, magnetic transitions~\cite{Rossel_1987}, and quantum criticality~\cite{Osato_2024}. YbCu$_4$In, in particular, undergoes a first-order valence transition around 42 K, as identified in previous studies~\cite{Nakamura_1990,Kojima_1990,Yamaoka_2023}. A noteworthy advancement is the recent discovery, via single-crystal X-ray diffraction experiments~\cite{Tsutsui_2016}, of electric quadrupoles emerging alongside the valence transition. This finding prompts a reassessment of existing valence studies on YbCu$_4T$, emphasizing the need to account for quadrupole effects.

YbCu$_4$Au is a candidate material exhibiting spin and valence fluctuations simultaneously. 
Polycrystalline YbCu$_4$Au exhibits a magnetic transition, as observed by specific heat, electrical resistivity, neutron diffraction, and nuclear magnetic resonance (NMR) measurements below 1.3 T~\cite{Rossel_1987,Bauer_1994,Yoshimura_2001,Yamamoto_2007,Wada_2008,Bauer_1997}. 
Moreover, studies on NMR above 2 T have indicated alterations in the electric field gradient (EFG) at the Cu site, potentially indicating a valence crossover~\cite{Wada_2008}. However, previous studies have also suggested an increased the possibility of an electric quadrupole contribution above 1.3 T, indicating a complex scenario~\cite{Takeuchi_2015}. Future experimental work focused on determining the valence of Yb will be crucial for clarifying these ambiguities.

\begin{figure}[h]
\begin{center}
\includegraphics[width=8.5cm,clip]{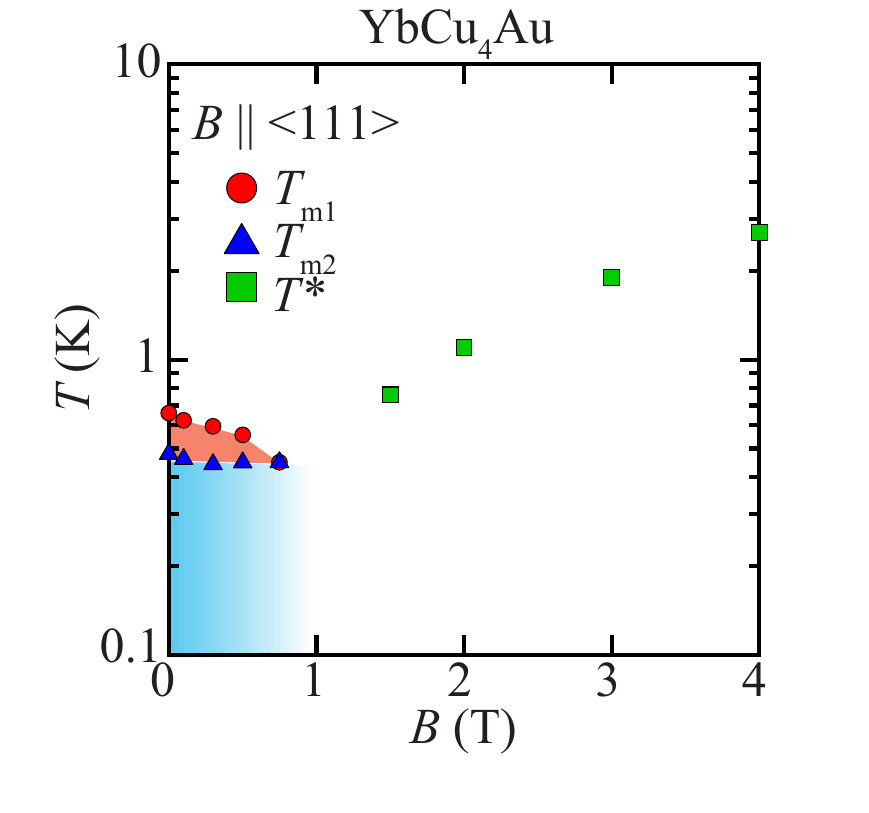}
\end{center}
\caption{(color online) $B$$\--$$T$ phase diagram of single-crystal YbCu$_4$Au for $B \parallel \left\langle 111 \right\rangle$. Phase transition and characteristic temperatures ($T_{\mathrm{m1}}$, $T_{\mathrm{m2}}$, and $T$*) are determined from specific heat measurements. The red and blue phases are magnetic.
}
\label{Fig.phase}
\end{figure}

In this study, we successfully grew a single crystal of YbCu$_4$Au and performed several experiments to investigate the magnetic field-induced criticality in this material. To the best of our knowledge, there have been no preports of growing single crystals of YbCu$_4$Au. 
We observed the crystal structure that reported a previous study~\cite{Adroja_1987}.
Our main results are represented by the phase diagram of single crystal YbCu$_4$Au for $B \parallel \left\langle 111 \right\rangle$ in Fig.~\ref{Fig.phase}, the details of which are discussed in Sec. III.
Our investigation revealed successive magnetic transitions below 1 T, and the valence of Yb above 2 T remained constant while the EFG changed. This indicates that the magnetic field above 2 T induces an electric quadrupole, which changes the EFG. We interpreted this phenomenon on the basis of the second-order perturbation theory, considering the crystal field and Zeeman effect. Our $B$$\--$$T$ phase diagram demonstrates the existence of a bicritical point near 1 T; a significant finding from our experimentals.

\section{\label{sec:level2}Experimental Details}
Single-crystal YbCu$_4$Au, approximately 10 mm$^3$ in size, was synthesized through the following process. Ingots of Yb (99.9\% purity), Cu powder (99.99\% purity), and Au powder (99.99\% purity) were combined in a Yb:Cu:Au molecular ratio of 1.1:4.0:1.0. This mixture was placed in an alumina crucible, which was then sealed with an alumina cap. The crucible was enclosed in quartz tubes under an Ar gas atmosphere and heated to 1373 K for 12 h. Subsequently, the tubes were cooled to 873 K at a rate of 1.5 K/h before being allowed to naturally reach room temperature.
For reference to YbCu$_4$Au, we also synthesized YbCu$_4$In.
For the growth of the single crystal YbCu$_4$In, we employed the flux method, using an In$\--$Cu flux as detailed in Ref.~\citen{Sarrao_1996}. 
High-purity materials (Yb, 99.9\% purity; Cu, 99.99\% purity; and In, 99.999\% purity) were placed in an alumina crucible and sealed within an evacuated quartz tube.

The phase purity of YbCu$_4$Au was examined using XRD and NPD measurements. Single crystals of YbCu$_4$Au were crushed into polycrystalline powder. Cu-K$\alpha$ radiation with a wavelength  of 1.54 \AA$ $ was used for XRD measurements. Additionally, neutron diffraction measurements were conducted using Ge(331) and Ge(311) reflections (2.20 and 2.01 \AA) on the HERMES~\cite{Nambu_2024} and AKANE~\cite{Ikeda_2024} diffractometers at Japan Research Reactor No. 3 (JRR-3), Japan. NPD measured at AKANE below 2 K was used for a $^3$He refrigerator.

The resistivity of a single crystal YbCu$_4$Au was measured using a four-probe method with an $ac$ resistance bridge in a standard configuration with a Quantum Design physical property measurement system (PPMS) above 2 K and a dilution refrigerator below 2 K. 
Above 2 K, we used a Quantum Design commercial superconducting quantum interference device (MPMS) to measure the $dc$ susceptibility of the single-crystal of YbCu$_4$Au.
The specific heat of the single-crystal of YbCu$_4$Au below 2 K was measured using the standard quasi-adiabatic heat-pulse method in a dilution refrigerator.
Above 2 K, we used a PPMS to measure the specific heat. 

We performed NMR measurements on the crystal used in the specific heat and electrical resistivity measurements. For NMR measurements, a crystal was shaped into a thin plate of $2 \times 1 \times 0.1$ mm$^3$ with the $\left\langle 111 \right\rangle$ direction normal to the plate. This operation enabled the radio frequency ($rf$) magnetic field, which is attenuated by the skin effect, to penetrate the interior of the sample. The NMR spectra were obtained by summing the Fourier transform of the spin-echo signal recorded at equally spaced magnetic fields with fixed $rf$ frequency. The orientation of the crystal with respect to the magnetic field was precisely controlled using a double-axis goniometer.

$\mu$SR measurements were performed using a dilution refrigerator with the D1 $\mu$SR spectrometer at the Materials and Life Science Experimental Facility in J-PARC, Japan, and the spectrometer (CHRONUS) at Port 4 in the RIKEN-RAL Muon Facility at the Rutherford Appleton Laboratory in ISIS Neutron and Muon Source, UK. 
The YbCu$_4$Au powder used in XRD and NPD was also used for the $\mu$SR measurements.

XAS experiments at the Yb-L$_{\mathrm{I\hspace{-1.2pt}I\hspace{-1.2pt}I}}$ edge were performed on the BL39XU beamline at SPring-8~\cite{Kawamura_2009}. 
The powder of YbCu$_4$Au used in XRD, NPD, and $\mu$SR measurements, was pelletized for the XAS measurements. 
The reference sample, YbCu$_4$In, was a single crystal used for the XAS measurements.

\section{\label{sec:level3}RESULTS AND ANALYSIS}
\subsection{\label{sec:level4}Crystal Structure}
\begin{figure}[h]
\vspace*{10pt}
\begin{center}
\includegraphics[width=7.5cm,clip]{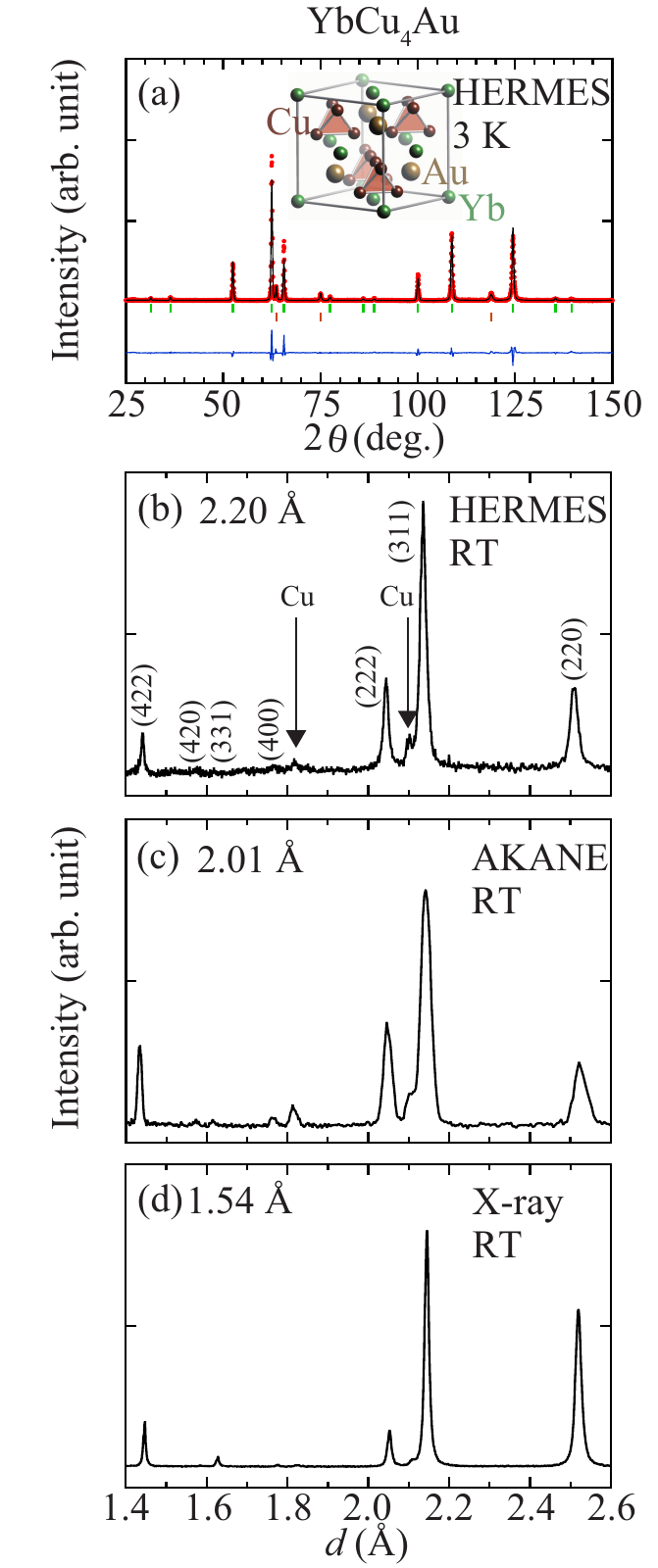}
\end{center}
\caption{(color online)(a) NPD profiles of YbCu$_4$Au refined using cubic MgCu$_4$Sn-type structure. The red circles, black line, blue line, green bars, and brown bars represent the experimental values, calculated values, the difference between experimental and calculated values, the Bragg positions of YbCu$_4$Au, and the Bragg positions of Cu, respectively. Inset represents the crystal structure of YbCu$_4$Au. NPD patterns with (b) 2.20 \AA$  $ and (c) 2.01 \AA, and (d) XRD pattern of YbCu$_4$Au. To compare NPD and XRD with different wavelengths, the horizontal axis was set using $d$, which is from the Bragg equation: $\lambda_a=2d\sin \theta$. Here, $\lambda_a$ represents the wavelength, with NPD wavelengths being 2.20 and 2.01 \AA, and the XRD wavelength being 1.54 \AA.
}
\label{Fig.1}
\end{figure}

Figures~\ref{Fig.1}(a)$\--$\ref{Fig.1}(d) show NPD and XRD patterns, respectively. 
All peaks can be assigned to the F$\bar{4}$3m (No. 216) symmetry of the cubic MgCu$_4$Sn-type phase. 
The crystal structure was successfully resolved using Rietveld analysis with the FULLPROF package~\cite{Rodriguez_1993}. 
These results are in agreement with previous studies~\cite{Rossel_1987,Bauer_1997,Sarrao_1999,Yoshimura_2001,Giovannini_2005,Giovannini_2008,Giovannini_2015}.

\subsection{\label{sec:level5}General Behaviors}
\begin{figure}[h]
\vspace*{10pt}
\begin{center}
\includegraphics[width=7.0cm,clip]{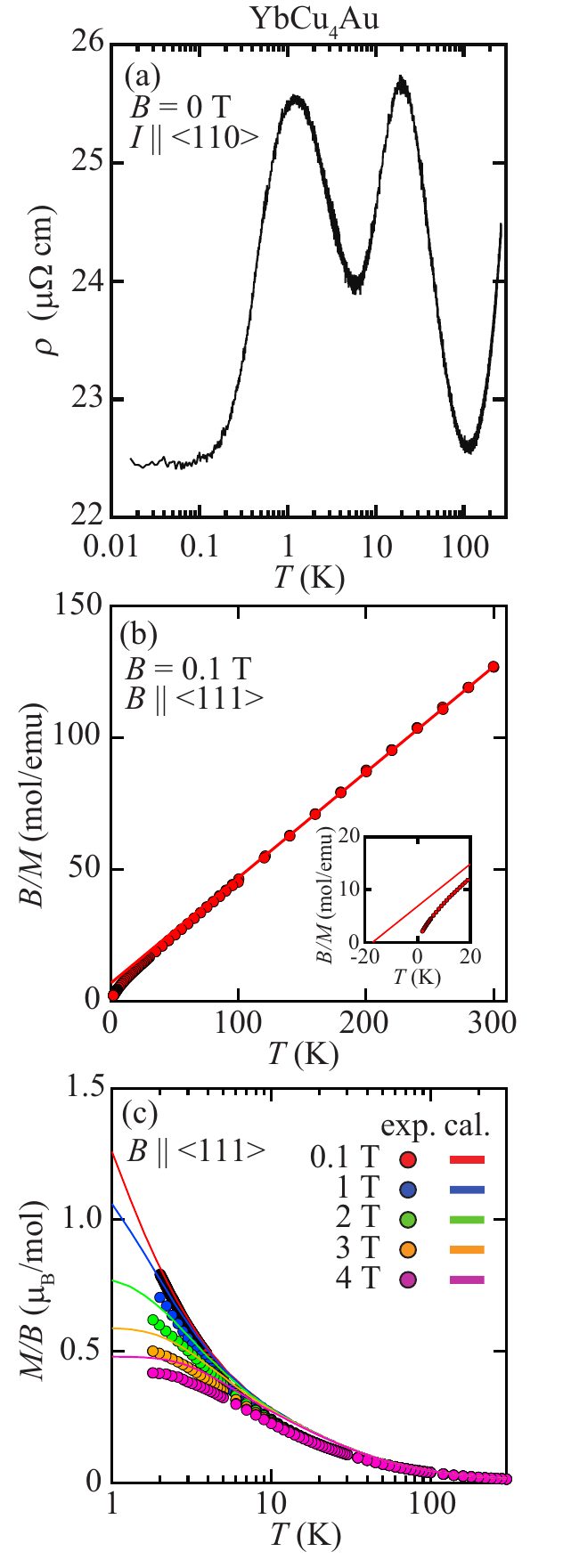}
\end{center}
\caption{(color online) Temperature dependence of (a) the resistivity, (b) inverse magnetic susceptibility for $B \parallel \left\langle 111 \right\rangle$, and (c) magnetic susceptibility for $B \parallel \left\langle 111 \right\rangle$ of YbCu$_4$Au. Inset represents the detail of inverse magnetic susceptibility between -20 and 20 K.
}
\label{Fig.2}
\end{figure}
Figure~\ref{Fig.2}(a) shows the temperature dependence of the resistivity of single crystal YbCu$_4$Au. 
The presence of two maxima at approximately 20 and 1 K indicates the behavior observed when Kondo effect and crystal field share the same energy scale~\cite{Bauer_1994}. 
In this case, the Kondo temperature was estimated to be between the two maxima, $T_\mathrm{K} \sim 6$ K. 
This behavior aligned with the findings from a previous study on polycrystalline YbCu$_4$Au~\cite{Bauer_1994}.

Figure~\ref{Fig.2}(b) shows the temperature dependence of the inverse susceptibility of single crystal YbCu$_4$Au at 1 T. 
Above 100 K, the inverse susceptibility shows a linear relationship with temperature. 
A Curie$\--$Weiss fitting was performed above 150 K, yielding an estimated Weiss temperature $T_{\rm w} \sim -17$ K [see inset of Fig.~\ref{Fig.2}(b)] and an effective magnetic moment $n_{\rm eff} = 4.46$ $\mu _{\rm B}$. 
Since the theoretical value of the effective magnetic moment of Yb$^{3+}$ is 4.54 $\mu _{\rm B}$, Yb$^{3+}$ primarily contributes to the magnetism of YbCu$_4$Au.

Figure~\ref{Fig.2}(c) shows the temperature dependence of the magnetic susceptibility for $B \parallel \left\langle 111 \right\rangle$. 
With increasing magnetic field, the magnetic susceptibility decreases below 10 K. 
For a quantitative understanding, we calculated the temperature dependence of the magnetic susceptibility using a mean-field approximation by diagonalizing the following Hamiltonian within the $J = \frac{7}{2}$ multiplet of Yb$^{3+}$:
\begin{eqnarray}
\text{\ensuremath{\mathscr{H}}}=\text{\ensuremath{\mathscr{H}}}_{\rm CEF}+\text{\ensuremath{\mathscr{H}}}_{\rm Z}+\text{\ensuremath{\mathscr{H}}}_{\rm D}
\label{eq:H_para}.
\end{eqnarray}
$\text{\ensuremath{\mathscr{H}}}_{\rm CEF}$ is the crystal electronic field (CEF) potential with the $T_d$ point group symmetry:
\begin{align}
\text{\ensuremath{\mathscr{H}}}_{\rm CEF}=W\left\{ x\frac{O_{40}+5O_{44}}{F_{4}}+\left(1-\left|x\right|\right)\frac{O_{60}-21O_{66}}{F_{6}}\right\} \nonumber \\
\label{eq:H_CEF},
\end{align}
where $O_i$ represents Stevens parameters~\cite{Stevens_1952}. $F_4$ and $F_6$ are 60 and 1260, respectively. $\text{\ensuremath{\mathscr{H}}}_{\rm Z}$ is Zeeman effect.
\begin{eqnarray}
\text{\ensuremath{\mathscr{H}}}_{\rm Z}=-g_{J}\mu_{B}\boldsymbol{J}\cdotp\boldsymbol{B}_{\rm ext}
\label{eq:Zeeman},
\end{eqnarray}
where the Lande $g$-factor $g_J$ is given as $g_{J} = \frac{8}{7}$ for Yb$^{3+}$. $\text{\ensuremath{\mathscr{H}}}_{\rm D}$ is the mean-field Hamiltonian for the exchange between magnetic dipoles.
\begin{eqnarray}
\text{\ensuremath{\mathscr{H}}}_{\rm D}=-\lambda_{d}\left\langle \boldsymbol{J}\right\rangle \cdotp\boldsymbol{J}
\label{eq:dipole},
\end{eqnarray}
$\left\langle \boldsymbol{J}\right\rangle$ is the thermal average value of the dipole. 
The CEF parameters were estimated via inelastic neutron scattering experiments: $W = -0.225$ meV and $x = -0.945$~\cite{Severing_1990}. $\lambda _d$ can be estimated from the Weiss temperature of the susceptibility: $\lambda_{d}=\frac{3g_{J}^{2}T_{w}}{zn_{\rm eff}^{2}}$. $z$ represents the number of nearest neighbor atoms. The $\frac{M}{B} \equiv \frac{N_{A} g_{J} \mu_{B}\left\langle \boldsymbol{J}\right\rangle\cdotp\boldsymbol{B}_{\rm ext}}{B_{\rm ext}^2}$ is the result of the mean-field calculation shown in Fig.~\ref{Fig.2}(c) using the solid lines. 
The observed susceptibility could be reasonably well reproduced by the calculations above 10 K, with an appropriate choice of the CEF parameters. 
However, a small systematic deviation, which could be attributed to the Kondo effect, was observed. 
The interactions, which might contribute to this discrepancy between $f$ and conduction electrons, are not explicitly considered in Eq. (1), which might be contributing to this discrepancy.

\subsection{\label{sec:level6}Magnetic Ordering}
\begin{figure}[t]
\vspace*{10pt}
\begin{center}
\includegraphics[width=8.0cm,clip]{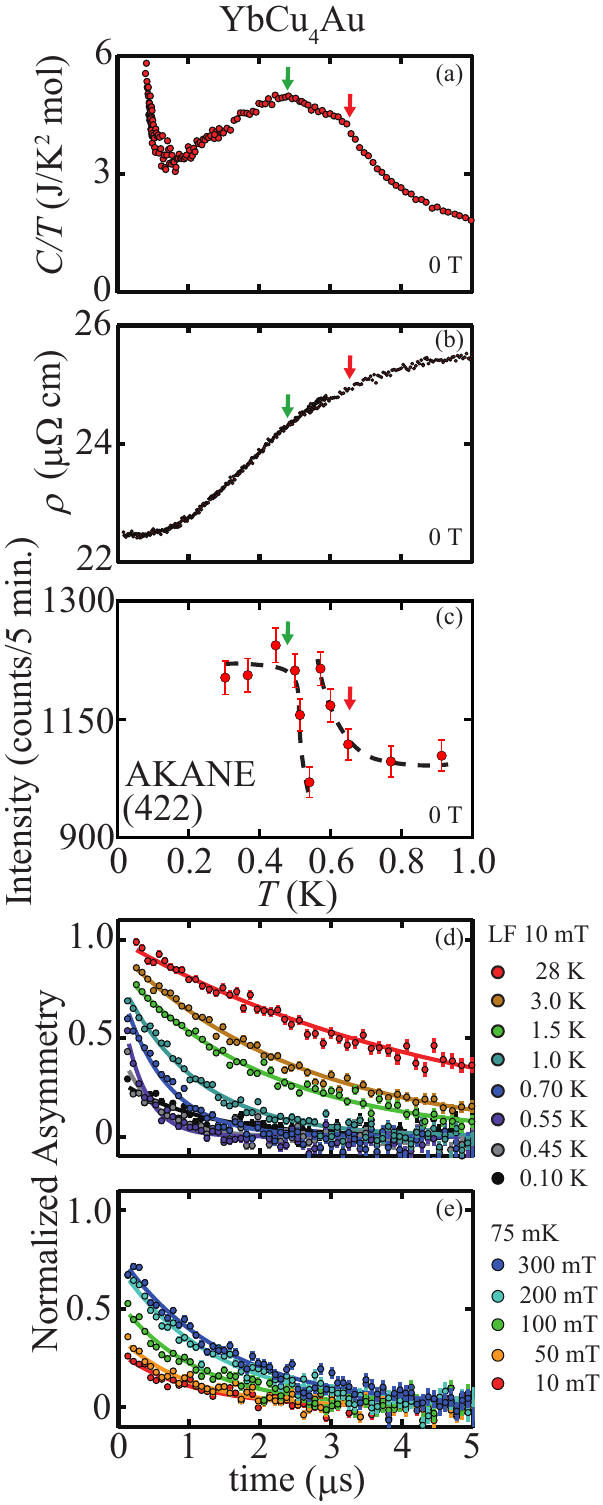}
\end{center}
\caption{(color online) Temperature dependence of (a) specific heat, (b) resistivity, (c) (422) peak intensity by NPD with AKANE under zero-fields of YbCu$_4$Au. The dashed lines are a guide for the eyes. (d) Temperature dependence of $\mu$SR spectra under longitudinal magnetic field 10 mT and (e) longitudinal magnetic field dependence of $\mu$SR spectra at 75 mK.
}
\label{Fig.3}
\end{figure}

\begin{figure}[t]
\vspace*{10pt}
\begin{center}
\includegraphics[width=8.5cm,clip]{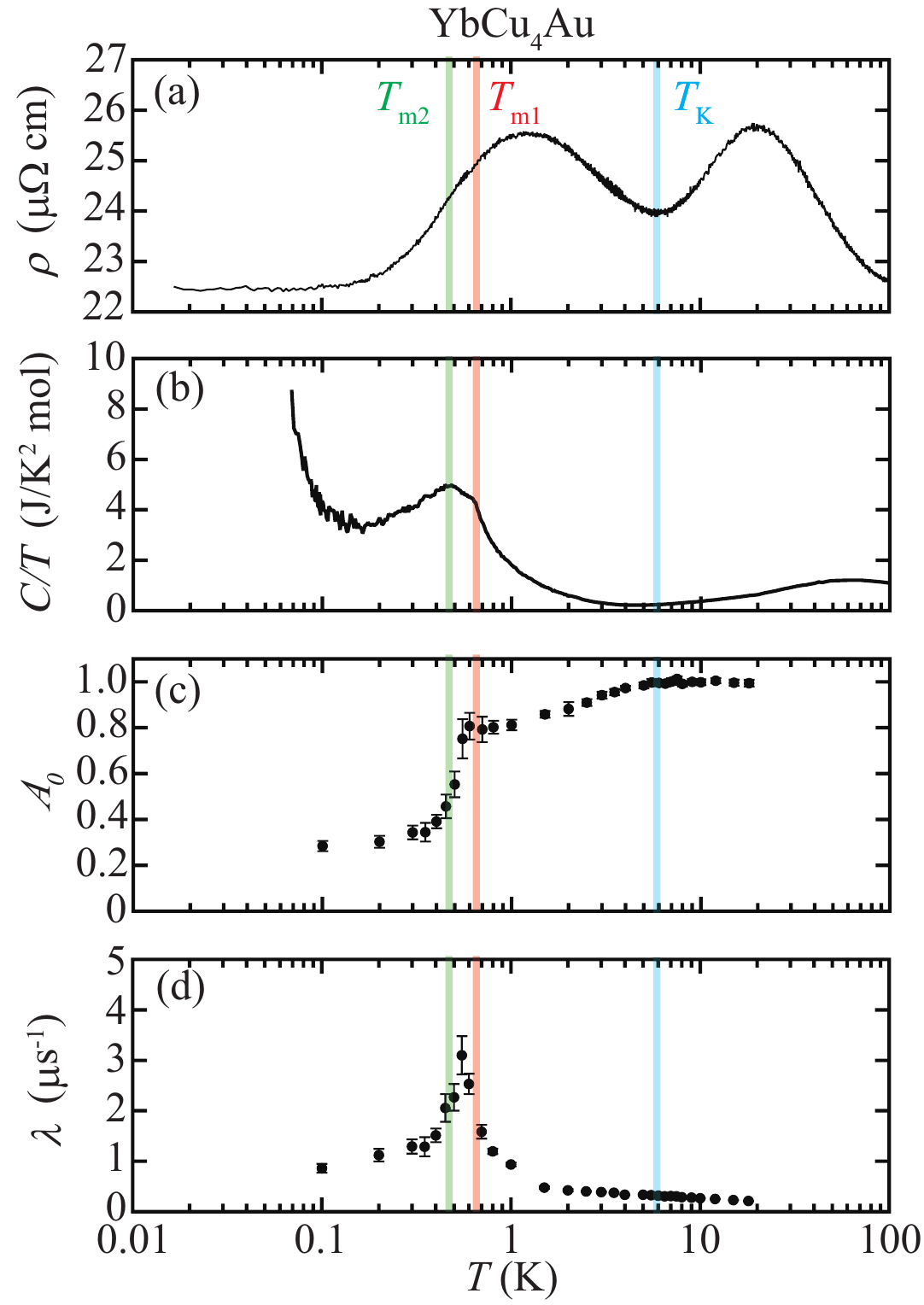}
\end{center}
\caption{(color online) Temperature dependence of (a) resistivity, (b) specific heat, (c) asymmetry, and (d) muon spin relaxation rate of YbCu$_4$Au. Bule, red, and green lines represent Kondo temperature, first-, and second-magnetic transition temperatures, respectively.
}
\label{Fig.4}
\end{figure}

In this section, we discuss the finding of magnetic successive transitions below 1 T.
Figure~\ref{Fig.3}(a) shows the temperature dependence of specific heat at zero magnetic field. 
The enhancement of specific heat below 0.2 K is attributed to the nuclear contribution. 
The anomalies related to the phase transition could be observed at $T_{\mathrm{m1}} = 0.66$ K and $T_{\mathrm{m2}} = 0.48$ K, as indicated by the red and green arrows in Fig.~\ref{Fig.3}(a), respectively. 
As shown in Fig.~\ref{Fig.3}(b), the slope also changes at the phase transition temperatures in resistivity, confirming consistency with the specific heat result. 
Figure~\ref{Fig.3}(c) shows the temperature dependence of the (422) signal intensity from neutron powder diffraction results shown in Fig.~\ref{Fig.1}(c). 
For reliability, we focused on the (422) signal for this study because (i) there are no impurity signals nearby, (ii) it is sharp, and (iii) it has high NPD signal intensity, as measured by AKANE.
Signal intensity decreases with increasing temperature in $T_{\mathrm{m1}}<T<T_{\mathrm{m2}}$. 
Since neutron diffraction experiment is a powerful tool for detecting both structural and magnetic transitions~\cite{Squires_2012}, this variation is due to either of the following: 1) a change in the form factor due to a structural phase transition, or 2) magnetic reflection suppressed by the development of another magnetic phase. 
Both possibilities show that the anomalies in specific heat and resistivity are not domain effects due to sample quality. 
To elucidate the phases, $\mu$SR measurements were performed, as shown in Figs.~\ref{Fig.3}(d) and \ref{Fig.3}(e). 
Below 28 K, the spectra exhibited an exponential-type shape,  indicating the presence of fast fluctuating internal fields at the muon site caused by surrounding electron spins. 
A slight recovery of the long-time region of the time spectrum was observed below 0.45 K, indicating the presence of static internal fields due to the magnetic order below 0.45 K, even though strong relaxation behaviors were observed. 
No distinct muon$\--$spin oscillations were observed even at 75 mK due to a loss in the initial asymmetry at $t=0$. 
Internal fields at the muon site are far beyond the time resolution of the detecting system for the pulsed muon beam at J-PARC. 
To estimate internal fields at the muon site,  measurements applying various longitudinal magnetic fields along the initial muon$\--$spin direction were performed at 75 mK, as shown in Fig.~\ref{Fig.3}(e). 
The decoupling behavior of the initial asymmetry was observed above approximately 100 mT or higher and was still half recovered even at 300 mT. 
This result suggests that internal fields at the muon stopping site exceed 300 mT~\cite{Hachitani_2006}.

To estimate the $\mu$SR parameters from the time spectra in shown Fig.~\ref{Fig.3}(c), the following equation was employed in our analysis:
\begin{eqnarray}
A\left(t\right)=A_{\rm0}e^{-\lambda t}
\label{eq:muon_para},
\end{eqnarray}
where $A_{\rm0}$ is the initial asymmetry of the $\mu$SR time spectrum at $t=0$. 
In general, the magnetic volume fraction in a magnetic ordered phase is 100\% when the asymmetry parameter at $t$, $A$($t$), is $\frac{1}{3}$ in the static limit case at $t$ = $\infty$~\cite{Umehara_1985}.
$\lambda$ is the muon$\--$spin relaxation rate. 
To discuss the electronic state of YbCu$_4$Au, the $\mu$SR results were compared with those of the resistivity and specific heat measurements, as shown in Fig.~\ref{Fig.4}. 
$A_{\rm0}$ decreased below $T_{\rm K} \sim 6$ K as estimated from the resistivity results, indicating that Kondo effect provides fluctuating internal fields at the muon site.
Anomalies related to phase transitions were observed below 1 K. 
These transitions were magnetic rather than structural because of the following reasons: (i) $\lambda$ increased above $T_{\rm m1}$, (ii) $A_{\rm0}$ decreased below $T_{\rm m1}$, and (iii) the recovery of $A$($t$) was observed in the long-time region below $T_{\rm m2}$. 
Notably, the two peaks of $\lambda$ caused by the critical slowdown behavior of electronic spin fluctuations in YbCu$_4$Au overlapped and attained the maximum in $\lambda$ between $T_{\rm m1}$ and $T_{\rm m2}$. 
These results show that the two anomalies found in the specific heat and electrical resistivity measurements were both from magnetic transitions.
In summary, the anomalies at $T_{\mathrm{m1}}$ and $T_{\mathrm{m2}}$ were derived from magnetic transitions rather than structural transitions and were not caused by domains of sample quality: (i) Anomalies were observed at $T_{\mathrm{m1}}$ and $T_{\mathrm{m2}}$ in specific heat and electrical resistivity measurements. (ii) NPD measurements indicated that the anomalies were not due to the domains of sample quality. (iii) $\mu$SR measurements showed that both transitions were magnetic. (iv) A change in NPD signal intensity was due to the magnetic reflection.

\subsection{\label{sec:level7}High Field State}
\begin{figure*}[t]
\vspace*{10pt}
\begin{center}
\includegraphics[width=16cm,clip]{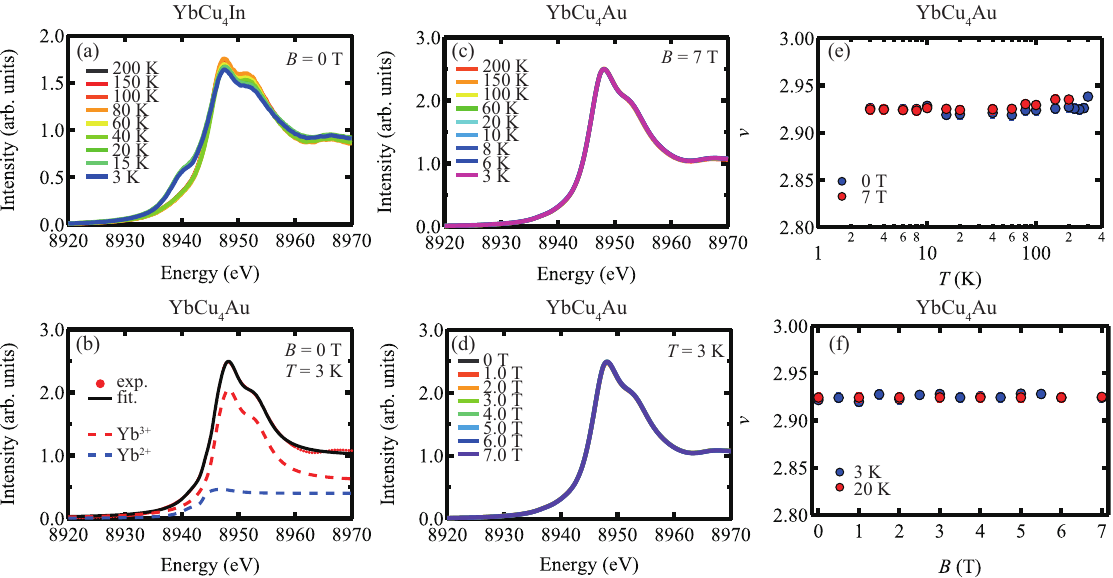}
\end{center}
\caption{(color online) (a) Temperature dependence of X-ray absorption spectra of YbCu$_4$In under zero-fields. (b) X-ray absorption spectrum of YbCu$_4$Au at 3 K. The red and blue dotted lines represent the calculated spectra from Yb$^{2+}$ and Yb$^{3+}$, respectively. (c) Temperature dependence of the X-ray absorption spectra of YbCu$_4$Au. (d) Magnetic field dependence of the X-ray absorption spectra at 3 K. (e) Temperature dependence and (f) magnetic field dependence of the average valence of Yb ion in YbCu$_4$Au. $\nu$ represents the average valence of Yb ion. 
}
\label{Fig.5}
\end{figure*}
In this section, we present the results of XAS measurements to determine the electronic states above 2 T. 
To determine the experimental conditions, we initially performed XAS measurements of YbCu$_4$In. Figure~\ref{Fig.5}(a) shows the temperature dependence of the Yb-L$_{\mathrm{I\hspace{-1.2pt}I\hspace{-1.2pt}I}}$ absorption spectra of YbCu$_4$In. 
As previously reported~\cite{Matsuda_2007}, YbCu$_4$In exhibits a first-order valence transition at 42 K, and this obvious difference gap around 8940 eV in Yb valence was clearly observed in our XAS measurements. 
The obtained spectral shape was consistent with the findings of the previous study~\cite{Matsuda_2007}, confirming the reliability of our experimental conditions. 
The Yb$^{3+}$ and Yb$^{2+}$ signals of Yb-L$_{\mathrm{I\hspace{-1.2pt}I\hspace{-1.2pt}I}}$ transitions appeared near 8950 and 8940 eV, respectively.

Regarding YbCu$_4$Au, the absorption spectrum at the Yb-L$_{\mathrm{I\hspace{-1.2pt}I\hspace{-1.2pt}I}}$ edge can be well described by Yb$^{3+}$ and Yb$^{2+}$ signals, as shown in Fig.~\ref{Fig.5}(b), where the dotted lines represent double Lorentzian functions. 
This observation suggests a coexistence of Yb valence states, namely Yb$^{2+}$ and Yb$^{3+}$, in YbCu$_4$Au.
The temperature dependence of the absorption spectrum at 7 T and the magnetic field dependence of the absorption spectrum at 3 K are shown in Figs.~\ref{Fig.5}(c) and \ref{Fig.5}(d), respectively. 
We derived the average Yb valence from these spectra, and its temperature and magnetic field dependences are shown in Figs.~\ref{Fig.5}(e) and \ref{Fig.5}(f), respectively. 
The average Yb valence was not 3.0, indicating a valence fluctuation state in YbCu$_4$Au. 
However, unlike the distinct changes observed in the case of YbCu$_4$In, no significant changes were observed in YbCu$_4$Au at various temperatures and magnetic fields. 
These results show that YbCu$_4$Au exhibits a valence fluctuation state in the temperature range $3\;\mathrm{K}\leq T\leq200\;\mathrm{K}$ and magnetic field range $0\;\mathrm{T}\leq B\leq7\;\mathrm{T}$, but without a distinct average value of valence change.

\subsection{\label{sec:level8}Phase Diagram}
\begin{figure}[h]
\vspace*{10pt}
\begin{center}
\includegraphics[width=8.5cm,clip]{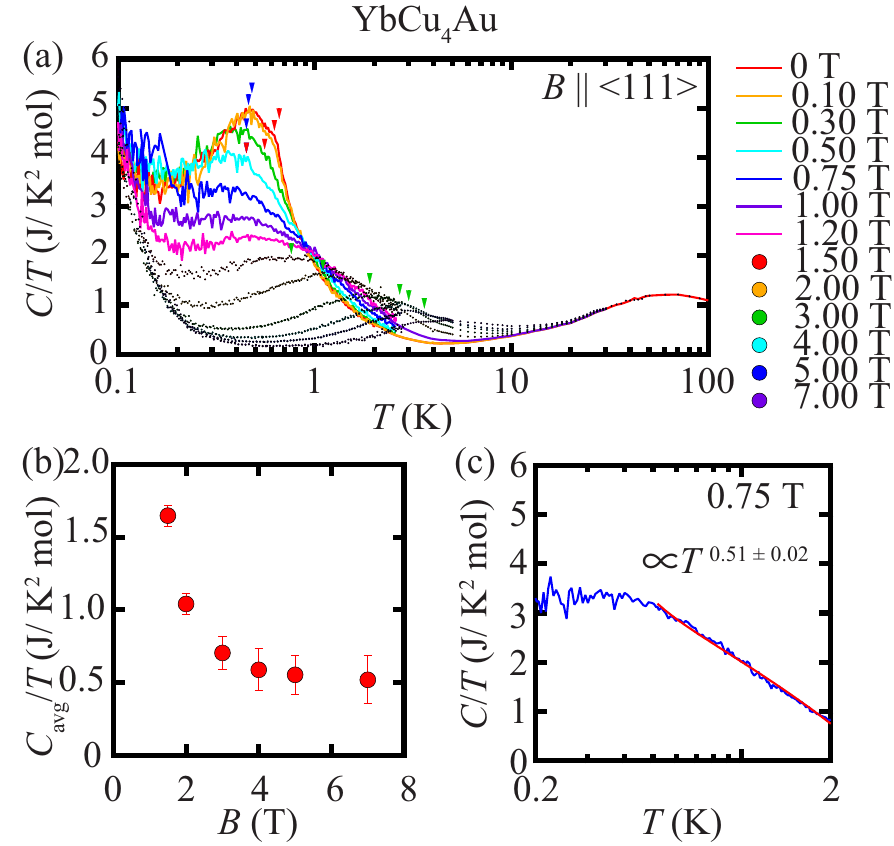}
\end{center}
\caption{(color online)  Temperature dependence of the specific heat ($\frac{C}{T}$) of YbCu$_4$Au for $B \parallel \left\langle 111 \right\rangle$. Red, blue, and green triangles represent $T_{\mathrm m1}$, $T_{\mathrm m2}$, and $T^*$, respectively. (b) Field dependence of $\frac{C}{T}$ averaged from 0.2 K to 0.3 K. (c) (blue) Temperature dependence of $\frac{C}{T}$ at 0.75 T and (red) fitting function: $\frac{C}{T}=a_{0}+a_{1}\left|T-T_{\mathrm{c}}\right|^{p}$ with $T_{\mathrm{c}} = 0.45$.
}
\label{Fig.6}
\end{figure}

Figure~\ref{Fig.6}(a) shows the temperature dependence of the specific heat ($\frac{C}{T}$) under various magnetic fields.
The significant increase in $\frac{C}{T}$ below 0.2 K is due to nuclear spins. 
Information about the nuclei contained in YbCu$_4$Au and the details of the contribution of nuclear spins are summarized in Supplementary Materials~\cite{Sup}.

Figure~\ref{Fig.6}(b) shows the field dependence of the average specific heat ($\frac{C_{\mathrm avg}}{T}$) in the temperature range between 0.2 and 0.3 K, where the contribution of nuclear specific heat is relatively small.
As reported from a previous study~\cite{Wada_2008}, below 1.3 T, an internal magnetic field originating from the magnetic phase is generated. 
To ignore the internal field, we focus on $\frac{C_{\mathrm avg}}{T}$ above 1.5 T.
Above 1.5 T, $\frac{C_{\mathrm avg}}{T}$ decreases dramatically with increasing magnetic field, indicating that a critical point exists below 1.5 T.
To investigate the contribution of electron spin fluctuations near the critical point, we analyzed the temperature dependence of the specific heat at 0.75 T, where the transition temperatures merged, using the Self-Consistent Renormalization (SCR) theory~\cite{Moriya_1985}, as shown in Fig.~\ref{Fig.6}(c). 
The SCR theory, developed by Moriya's group, successfully explains the critical phenomena in itinerant magnetic systems. 
Now, by analyzing the experimental data on the basis of SCR theory, we are able to determine the type of spin fluctuations.
As indicated by the red line, the experimental results were well reproduced by $\frac{C}{T}\propto a_{0}-a_{1}T^{p}$ ($a_0 =3.82 \pm 0.08$, $a_1 = 2.46 \pm 0.08$, $p=0.51 \pm 0.02$). 
Based on the SCR theory, these fitting values suggest that the three-dimensional antiferromagnetic fluctuations mainly contribute. 
This result is consistent with the previous NMR result: $\frac{1}{T_{1}}\propto T^{\frac{3}{4}}$.~\cite{Wada_2008, Yamamoto_2007}
Despite the coexistence of two magnetic transitions, only a single fluctuation was observed, indicating that the other fluctuation exhibited a significant magnetic anisotropy. 
In previous studies on YbCu$_{4.4}$Au$_{0.6}$, ferromagnetic fluctuations were observed at zero fields by $\mu$SR measurement~\cite{Carretta_2009}.
Ferromagnetic fluctuations with high uniaxiality could have been suppressed along the $B \parallel \left\langle 111 \right\rangle$ direction. 
Summarizing these points, (i) successive transition temperatures merge around 1 T, (ii) $\frac{C_{\mathrm avg}}{T}$ diverges towards lower fields above 1.5 T, and (iii) the temperature dependence of the specific heat at 0.75 T is reproduced by the SCR theory. 
Therefore, we concluded that the critical point existed near 1 T.

We made the $B$$\--$$T$ phase diagram of YbCu$_4$Au based on the results of the specific heat measurements, as shown in Fig.~\ref{Fig.phase}. 
The peaks observed above 1.5 T and below 10 K can be attributed to the crystal field ground doublet split by the Zeeman effect. 
The temperature corresponding to the apex of these peaks is denoted as $T$*. Figure~\ref{Fig.phase} shows the $B$$\--$$T$ phase diagram of single crystal YbCu$_4$Au, featuring $T_{\mathrm{m1}}$, $T_{\mathrm{m2}}$, and $T$*. Notably, in the vicinity of 1 T, successive transition lines merge.
Assuming that the paramagnetic phase at zero field undergoes a second-order instability, transitioning into the intermediate phase (red in Fig.~\ref{Fig.phase}), as we decrease the temperature at a fixed field.
This phase subsequently undergoes a second-order instability toward the low-temperature phase (blue in Fig.~\ref{Fig.phase}) at a further lower temperature. 
Based on thermodynamics, these phases cannot meet at a finite angle, i.e., Fig.~\ref{Fig.phase} is impossible~\cite{Yip_1991}. As shown in Fig.~\ref{Fig.3}(c), the signal intensity of (422) changes continuously at $0.6\;\mathrm{K}<T<1\;\mathrm{K}$ , indicating that the phase transition between the paramagnetic and the intermediate phases is of the second order. 
Conversely, as the signal intensity of (422) shows a discontinuous change near 0.6 K, the phase transition between the intermediate and low-temperature phases is of the first order. Furthermore, as shown in Fig.~\ref{Fig.6}(b), the specific heat decreases continuously with increasing magnetic field below 0.3 K. Therefore, the change in electronic state below 0.3 K with increasing magnetic field represents a crossover. 
These findings can be summarized as follows: (i) the transition between the paramagnetic and intermediate phases is of the second order, (ii) the transition between intermediate and low-temperature phases is of the first order, and (iii) a crossover is observed from a low-temperature phase to the phase above 2 T. Thermodynamics predicts the appearance of a bicritical point in this situation near 1 T~\cite{Yip_1991}. 

\section{\label{sec:level9}Discussion}
We demonstrate that the observed temperature-dependent change in the EFG, as reported in previous NMR studies~\cite{Wada_2008}, can be comprehended in the context of magnetic field-induced quadrupoles. 
The crystal field ground state of YbCu$_4$Au is a $\Gamma_7$ doublet, with the first excited state being a $\Gamma_8$ quartet, and the second excited state a $\Gamma_6$ doublet~\cite{Severing_1990}.
Under our experimental conditions, the effect of the external magnetic field can be treated as a second-order perturbation. 
In this case, as the $\Gamma_7$ doublet mixes with the excited state, the wavefunction can be expressed as follows:
\begin{align}
\left\langle \Gamma_{7}^{i*}\right| & =\left\langle \Gamma_{7}^{i}\right|+\sum_{k=x,y,z}\sum_{p,q}\left\{ a_{6}\left(\left\langle \Gamma_{7}^{i}\left|J_{k}B_{k}\right|\Gamma_{6}^{q}\right\rangle \left\langle \Gamma_{6}^{q}\right|\right)\right.\nonumber \\
 & \left.+a_{8}\left(\left\langle \Gamma_{7}^{i}\left|J_{k}B_{k}\right|\Gamma_{8}^{p}\right\rangle \left\langle \Gamma_{8}^{p}\right|\right)\right\}, 
\end{align}
where $a_{r}=\frac{g_{J}\mu_{B}}{\triangle_{r}}$ and $\triangle_{r}$ is the energy level difference between the ground and excited states. 
The wavefunctions for each of the doublet and quartet states are indexed by $i$, $p$, and $q$.
Using this mixed wave function, the expectation values of the five quadrupoles can be estimated as follows: $O_{20} = 3J_z^2 - J\left(J+1\right)$, $O_{22} = J_x^2 - J_y^2$, $O_{xy} = J_xJ_y + J_yJ_x$, $O_{yz} = J_yJ_z + J_zJ_y$, and $O_{zx} = J_zJ_x + J_xJ_z$. 
With the magnetic field direction represented in polar coordinates, the  value of $O_{20}$ in the polycrystalline case of YbCu$_4$Au can be obtained as follows:
\begin{eqnarray}
\left\langle O_{20}\right\rangle^i =\int\int\sin\theta d\theta d\phi\left\langle \Gamma_{7}^{i*}\left|O_{20}\right|\Gamma_{7}^{i*}\right\rangle \neq0.
\end{eqnarray}
The same calculation also provides finite expectation value for the other quadrupoles. 
Therefore, quadrupoles can be detected in the polycrystalline YbCu$_4$Au. 
As reported for PrTi$_2$Al$_{20}$~\cite{Taniguchi_2016,Taniguchi_2019}, NMR is a powerful tool for observing EFG changes attributed to the quadrupoles.
In this model, quadrupoles are generated by Zeeman effect.
Therefore, our experimental results indicate that the bicritical point near 1 T arises from the competition between RKKY interactions, providing the magnetic phases, and Zeeman effect.

The interesting and potentially exotic is bicriticality in YbCu$_4$Au involving itinerant antiferromagnetism and ferromagnetism.
This bicriticality can lead to unconventional phenomena, as described in Ref.~\citen{Andrade_2014}, although it has not yet been theoretically clarified. 
A study on the bicriticality between itinerant antiferromagnetism and ferromagnetism has been reported on YbRh$_2$Si$_2$. 
By applying pressure or doping with Co, successive transitions appear~\cite{Lausberg_2013}, and the bicriticality is exhibited near the ambient pressure in YbRh$_2$Si$_2$. 
Due to the bicriticality, it has been suggested that a small amount of Ir-doped Yb(Rh$_{0.94}$Ir$_{0.06}$)$_2$Si$_2$ exhibits a spin liquid state, making it a topic of ongoing research~\cite{Lausberg_2013, Friedemann_2009}. 
Since Yb(Rh$_{1-x}$Co$_x$)$_2$Si$_2$ exhibits an intermediate antiferromagnetic phase and a low-temperature ferromagnetic phase, its phase diagram is the reverse of that of YbCu$_4$Au. 
The ferromagnetism of YbCu$_4$Au in the intermediate phase may be due to the magnetic geometrical frustration, as observed in YbCu$_4$Ni~\cite{Sereni_2018}, because the point group of the Yb site in YbCu$_4$Au is $T_{\rm d}$ [see inset of Fig. 2(a)]. Therefore, the results of this study provide valuable insights for research on bicriticality, particularly in comparison with YbRh$_2$Si$_2$.

\begin{figure}[h]
\vspace*{10pt}
\begin{center}
\includegraphics[width=8.5cm,clip]{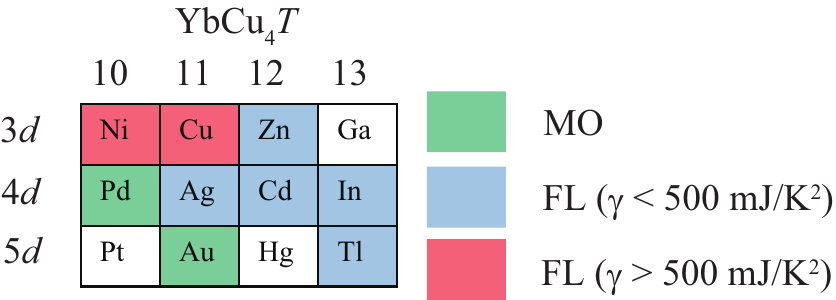}
\end{center}
\caption{(color online) Periodic table showing transition metal elements for which cubic YbCu$_4$$T$ ($T$: transition metal) can be synthesized. White block represents an element for which there have been no reports of cubic YbCu$_4$$T$. Green, blue, and red blocks represent elements that exhibit a magnetic transition, Fermi liquid with specific heat coefficient $\gamma<500$ mJ/K$^2$, and Fermi liquid with $\gamma>500$ mJ/K$^2$ in cubic YbCu$_4$$T$, respectively. This figure shows the data presented by YbCu$_4$$T$ ($T$ = Ni~\cite{Sereni_2018}, Cu~\cite{Tsujii_1997}, Zn~\cite{Sarrao_1999}, Pd~\cite{Rossel_1987}, Ag~\cite{Lawrence_2001}, Cd~\cite{Sarrao_1999}, In~\cite{Lawrence_2001}, and Tl~\cite{Lawrence_2001}).
}
\label{Fig.7}
\end{figure}
Finally, we discuss the electronic state of YbCu$_4$$T$. 
Figure~\ref{Fig.7} shows the periodic table for YbCu$_4$$T$ to compare the relationship between transition metals and their electronic states. 
The higher the group number in the periodic table, the stronger Kondo effect, leading to a Fermi liquid state~\cite{Sarrao_1999, Lawrence_2001}. 
Conversely, a smaller group number implies stronger RKKY interaction, resulting in magnetic ordering~\cite{Rossel_1987}. 
These results suggest that the group number can elucidate the Doniach phase diagram~\cite{Doniach_1977}. 
However, in cases such as YbCu$_4$Au and YbCu$_4$Zn, the atomic number alone cannot account for the Doniach phase diagram. 
This consideration indicates that the transition metal electrons are essential for modifying the Fermi surface. 
Therefore, the quantum criticality of YbCu$_4$$T$ can potentially be explained by considering external parameters (such as magnetic field and external pressure) for a single material while disregarding transition metal effects. 
In essence, the study of magnetic field and external pressure effects using single crystals of YbCu$_4$Au from this work and YbCu$_4$Ni~\cite{Osato_2024, Sereni_2018}, which exhibits quantum criticality at ambient pressure and zero magnetic field, will provide important insights.

\section{\label{sec:level10}Conclusion}
In this study, we synthesized single crystals of YbCu$_4$Au and performed a comprehensive analysis using various techniques including XRD, NPD, NMR, resistivity, magnetization, specific heat, $\mu$SR, and XAS measurements. 
Our investigations below 1 T, particularly the specific heat, resistivity, and $\mu$SR measurements at $T_{m1} = 0.66$ K and $T_{m2} = 0.48$ K, revealed magnetic successive transitions. 
Additionally, no significant change in the Yb valence was observed in the XAS measurements below 7 T and above 3 K. 
Based on these findings, we inferred that the quadrupole is induced by the magnetic field above 2 T. 
Notably, our specific heat measurements under several magnetic fields revealed the presence of a bicritical point near 1 T. 
Therefore, around the bicritical point, there is competition between RKKY interactions, providing the magnetic phases, and the Zeeman effect.
These results offer valuable insights into the magnetic and electronic behaviors of YbCu$_4$Au, contributing to a further understanding of its unique characteristics.

\section*{Acknowledgment}
\begin{acknowledgment}


We thank K. Ishida, S. Kitagawa, S. Tsutsui, and Y. Shimura for stimulating discussions, and M. Ohkawara for his technical supports at HERMES and AKANE. The neutron diffraction experiment at JRR-3 was carried out under the general user program managed by The Institute for Solid State Physics, the University of Tokyo (Proposal Nos. 22410, 22616, 23409, and 23608), and supported by the Center of Neutron Science for Advanced Materials, Institute for Materials Research, Tohoku University. The $\mu$SR measurements at the Materials and Life Science Experimental Facility of J-PARC (Proposal Nos. 2020B0402, 2021B0131) and RIKEN-RAL Muon Facility in the Rutherford Appleton Laboratory (Experiment No. RB2070002), and the XAS measurement at Spring-8 (Proposal No. 2022A1257), and NMR measurements at high field laboratory, Institute for Materials Research, Tohoku University (Proposal Nos. 202112-HMKPB-0404 and 202212-HMKPB-0404), were performed under user program. This work was financially supported by JSPS/MEXT Grants-in-Aids for Scientific Research (Grant Nos. 19K23417, 21K13870, and 23K13051).

\end{acknowledgment}





\bibliography{YbCu4Au}

\end{document}